\documentclass[lettersize,journal]{IEEEtran}
\usepackage{amsmath,amsfonts}
\usepackage{algorithmic}
\usepackage{array}
\usepackage[caption=false,font=footnotesize,labelfont=rm,textfont=rm]{subfig}
\usepackage{textcomp}
\usepackage{stfloats}
\usepackage{url}
\usepackage{verbatim}
\usepackage{graphicx}
\usepackage{amsmath}
\usepackage{bm}
\usepackage{amssymb}
\usepackage{algorithmic}
\usepackage{algorithm}
\usepackage{amsthm}
\usepackage{color}
\usepackage{hyperref}
\allowdisplaybreaks[4]
\def\BibTeX{{\rm B\kern-.05em{\sc i\kern-.025em b}\kern-.08em
    T\kern-.1667em\lower.7ex\hbox{E}\kern-.125emX}}
\usepackage{balance}

\begin{document}
\newcommand{\wcm}{\bm{w}_{\text{c},m}}
\newcommand{\Wcm}{\mathbf{W}_{\text{c},m}}
\newcommand{\Rsm}{\mathbf{R}_{\text{s},m}}
\newcommand{\Rxm}{\mathbf{R}_{x,m}}
\newcommand{\pmax}{P_\text{max}}
\newcommand{\Nt}{N_\text{t}}
\newcommand{\Nr}{N_\text{r}}
\newcommand{\dmin}{d_\text{min}}
\newcommand{\vmax}{v_\text{max}}
\newcommand{\rank}{\text{rank}}
\newcommand{\rhoml}{\rho_{m,l}}

\newcommand{\Gammal}{\Gamma_l}
\newcommand{\coverageArea}{\mathcal{A}_\text{e}}
\newcommand{\timeIndex}{t_m}
\newcommand{\alphaMK}{\alpha_{m,k}}
\newcommand{\Aml}{\mathbf{A}_{m,l}}
\newcommand{\Zmk}{\mathbf{Z}_{m,k}}
\newcommand{\Gml}{\tilde{G}_{m,l}}
\newcommand{\RankC}{R_\text{c}}
\newcommand{\lambdaMR}{\lambda_{m}^r}
\newcommand{\uMR}{\bm{u}_{m,r}}
\newcommand{\bmrn}{b_{m,n}^r}
\newcommand{\bmrnHat}{b_{m,\hat{n}}^r}
\newcommand{\phimrn}{\phi_{m,n}^r}
\newcommand{\phimrnHat}{\phi_{m,\hat{n}}^r}
\newcommand{\deltaPhi}{\Delta \phi_{m,n,\hat{n}}^r}
\newcommand{\GlLowerBound}{f_{l,m,n,\hat{n},r}^{\rm lb}}
\newcommand{\Iml}{\tilde{I}_{m,l}}
\newcommand{\RankS}{R_\text{s}}
\newcommand{\lambdaMRforI}{\tilde{\lambda}_{m}^{\tilde{r}}}
\newcommand{\uMRforI}{\tilde{\bm{u}}_{m,{\tilde{r}}}}
\newcommand{\bmrnForI}{\tilde{b}_{m,n}^{\tilde{r}}}
\newcommand{\phimrnForI}{\tilde{\phi}_{m,n}^{\tilde{r}}}
\newcommand{\bmrnHatForI}{\tilde{b}_{m,\hat{n}}^{\tilde{r}}}
\newcommand{\phimrnHatForI}{\tilde{\phi}_{m,\hat{n}}^{\tilde{r}}}
\newcommand{\deltaPhiForI}{\Delta \tilde{\phi}_{m,n,\hat{n}}^{\tilde{r}}}
\newcommand{\IlUpperBound}{f_{l,m,n,\hat{n},\tilde{r}}^{\rm ub}}
\newcommand{\vml}{\tilde{\bm{v}}_{m,l}}
\newcommand{\deltaQ}{\Delta \bm{q}_{m,n,\hat{n}}}
\newcommand{\deltaQPrev}{\Delta \bm{q}_{m,n,\hat{n}}^{[i-1]}}

\newcommand{\cosbraces}[1]{\cos \left\{ #1 \right\}}
\newcommand{\sinbraces}[1]{\sin \left\{ #1 \right\}}
\newcommand{\braces}[1]{\left( #1 \right)}
\newcommand{\midBraces}[1]{\left[ #1 \right]}
\newcommand{\bigBraces}[1]{\left\{ #1 \right\}}
\newcommand{\VertSquare}[1]{\Vert #1 \Vert^2}
\newcommand{\Ta}{\mathbf{T}_{\text{a},m}}
\newcommand{\Tb}{\mathbf{T}_{\text{b},m}}
\newcommand{\svTrans}{\bm{a}_{m}}
\newcommand{\svRec}{\bm{b}_k}
\newcommand{\amk}{\bm{a}_{m,k}}
\newcommand{\aml}{\tilde{\bm{a}}_{m,l}}
\newcommand{\bmk}{\bm{b}_{m,k}}
\newcommand{\pLEO}{\bm{p}_{\text{LEO},m}}
\newcommand{\pCEl}{\bm{p}_{\text{CE},l}}
\newcommand{\pSEk}{\bm{p}_{\text{SE},k}}
\newcommand{\qm}{\bm{q}_m}
\newcommand{\optimalQm}{\bm{q}_m^{(*)}}
\newcommand{\optimalOmega}{\omega^{(*)}} 
\newcommand{\qnm}{\bm{q}_{m,n}}
\newcommand{\qnhatm}{\bm{q}_{m,\hat{n}}}
\newcommand{\xnm}{x_{m,n}}
\newcommand{\ynm}{y_{m,n}}
\newcommand{\ThetaSEk}{\Theta_{\text{SE},k}}
\newcommand{\PhiSEk}{\Phi_{\text{SE},k}}
\newcommand{\Ys}{\mathbf{Y}_{\text{s},m,k}}
\newcommand{\ys}{\bm{y}_{\text{s},m,k}}
\newcommand{\yc}{y_{\text{c},m,l}}
\newcommand{\us}{\bm{u}_{\text{s},m,k}}
\newcommand{\X}{\mathbf{X}}
\newcommand{\Ns}{\mathbf{N}_{\text{s},m,k}}
\newcommand{\ns}{\bm{n}_{\text{s},m,k}}
\newcommand{\senNoisePower}{\sigma_k^2}
\newcommand{\commNoisePower}{\sigma_l^2}
\newcommand{\xiMK}{\bm{\xi}_{m,k}}
\newcommand{\Ximk}{\mathbf{\Xi}_{m,k}}
\newcommand{\phiK}{\bm{\phi}_k}
\newcommand{\bmAlphaMK}{\bm{\alpha}_{m,k}}
\newcommand{\SPEB}{{\rm SPEB}_{m,k}}
\newcommand{\partials}[2]{\frac{\partial #1}{\partial #2}}
\newcommand{\etal}{\emph{et al.}~}

\title{Joint Beamforming and Position Design for Movable Antenna Assisted LEO ISAC Systems}

\author{Hanfu Zhang, Erwu Liu

\thanks{
H. Zhang, and E. Liu are with the College of Electronic and Information Engineering, Tongji University, Shanghai 201804, China. E. Liu is also with the Shanghai Institute of Intelligent Science and Technology, Tongji University, Shanghai 201804, China. (e-mail: hanfuzhang@tongji.edu.cn; erwu.liu@ieee.org).

\emph{Corresponding author: Erwu Liu.}
}}
\maketitle
\begin{abstract}
\emph{Low earth orbit} (LEO) satellite-assisted \emph{integrated sensing and communications} (ISAC) systems have been extensively studied to achieve ubiquitous connectivity. However, the severe signal attenuation and limited transmit power at LEO satellites can degrade ISAC performance. To address this issue, this paper investigated \emph{movable antenna} (MA)-assisted LEO ISAC systems. We derive the communication \emph{signal-to-interference-plus-noise ratio} (SINR) and the sensing \emph{squared position error bound} (SPEB) for evaluating the ISAC performance. Then, we jointly optimize the transmit beamforming and the MA positions to minimize the SPEB under the SINR constraints, total transmit power constraint, and several inherent physical constraints of the MA array. We first simplify the complex problem using the \emph{semidefinite relaxation} (SDR). Then, we present a novel \emph{alternating optimization} (AO)-based algorithm to decouple the original problem into two subproblems, consequently convexified and solved. Simulations demonstrate the convergence and effectiveness of the proposed algorithm. Better trade-off between communication and sensing performance, and at least $25\%$ gain in sensing performance are achieved, compared to the benchmarks.
\end{abstract}


\section{Introduction}
\label{section: Introduction}

\subsection{Motivation and Challenges}
\IEEEPARstart{T}{he} integration of sensing and communication capabilities into one network, i.e. \emph{integrated sensing and communications} (ISAC) \cite{9737357, 9606831}, has been recognized as one of the key scenarios for the future \emph{sixth-generation} (6G) mobile networks \cite{5307322}. ISAC improves the resource utilization and reduces the equipment size, the energy consumption, and the mutual interference between the sensing and communication signals effectively \cite{10012421}. Extensive research has been conducted on the information-theoretical limits \cite{9705498, 10147248}, waveform design \cite{10042425, 10054103}, receive signal processing \cite{10964081, 11114786}, etc., of the ISAC systems.

However, conventional terrestrial ISAC systems cannot meet the requirement of ubiquitous connectivity of 6G \cite{5307322} in certain scenarios, such as oceans or remote mountainous areas where the infrastructure cannot be deployed. To address this issue, the \emph{low earth orbit} (LEO) satellites have been considered to achieve seamless global coverage from space \cite{10851844, 10478820}. The LEO satellites orbit at an altitude of $200$ to $2,000$ kilometers above the earth's surface and can be launched at a relatively low cost \cite{10478820}. Therefore, the LEO satellite is also regarded as a key component of the 6G networks \cite{10851844}.

Due to the considerable distance between LEO satellites and the earth surface, signal propagation experiences significant latency and severe attenuation, which degrades both communication and sensing performance. Additionally, the LEO satellites typically have limited payload capabilities, which constrain the size and weight of ISAC hardware and thereby restrict the system's transmit power \cite{10574260}. To maximize power utilization and further enhance communication and sensing performance, we introduce an emerging technology, \emph{movable antennas} (MAs) \cite{10318061, 10243545, 10354003}, which can adjust the position of each transmit antenna and increase the design flexibility, to replace the traditional \emph{uniform planar arrays} (UPAs).

\subsection{Related Works}
The LEO satellites have been extensively considered to assist in both communication and sensing systems. For example, You \etal \cite{9110855} established the massive \emph{multiple-input multiple-output} (MIMO) channel model for LEO satellite communications. The average signal-to-leakage-plus-noise ratio and the average signal-to-interference-plus-noise ratio were maximized by developing precoding and receiver design algorithms. A space angle based \emph{communication equipment} (CE) grouping algorithm was also proposed for the metrics to reach the upper bounds. Zheng \etal \cite{9849035} introduced the emerging technology of \emph{intelligent reflecting surfaces} (IRSs) at both the LEO satellite side and the ground side, i.e., two-sided cooperative IRSs, to further enhance the system communication performance. The overall channel gain between the satellite and the CE on the ground was maximized by conducting joint active and passive beamforming. Specifically, the high-dimensional channel matrix was first decomposed into a simpler form. Consequently, the original non-convex problem was decoupled into two subproblems at the satellite side and the ground side, respectively. Moreover, the channel estimation and beam tracking were also considered by proposing a transmission protocol.

On the other hand, the LEO satellites have also shown the potential to improve sensing performance. Emenonye \etal \cite{10992253} theoretically demonstrates the feasibility of using LEO satellites for nine-dimensional localization, including the estimation of the \emph{three-dimensional} (3D) position, 3D orientation, and 3D velocity. The fundamental limits were analyzed by deriving \emph{Fisher's information matrix} (FIM). The system parameters such as the number of LEO satellites and receiving antennas for achieving different levels of localization accuracy were also determined. Zhu \etal \cite{10695037} further improved the accuracy of the massive MIMO LEO localization systems at the practical level. The FIM for 3D localization was derived, and the sensing metric \emph{squared position error bound} (SPEB) was calculated based on the FIM. A worst-case sum SPEB minimization problem was consequently formulated when the prior position knowledge was imperfect. A codebook-based robust precoding scheme was proposed by adopting the \emph{majorization-minimization} (MM) method.

By jointly considering communication and sensing, some works have also used the LEO satellites to assist ISAC systems. For example, You \etal \cite{10478820} developed a wideband massive MIMO LEO system to support communication and localization simultaneously. The localization performance bound, SPEB, and the communication metric, \emph{spectral efficiency} (EE), were derived. A multi-objective problem was then formulated to improve both performance. You \etal \cite{9852292} also considered beam squint effects and characterized the statistical wave propagation properties in the massive MIMO LEO systems. Based on the analysis, a beam squint-aware ISAC problem was formulated to maximize the EE of the communications while maintaining the sensing performance. The Dinkelbach's algorithm was utilized to transform the problem into several subproblems. Consequently, the subproblems were solved alternatively via the Lagrangian dual transform, the quadratic transform, and the Lagrange multiplier method. The simulation results verified that the proposed algorithm mitigated the beam-squint effects.

Park \etal \cite{11143190} investigated a bistatic ISAC framework to mitigate severe radar echo loss, and employed the \emph{rate-splitting multiple access} (RSMA) approach in LEO ISAC systems for robust and effective interference management. The RSMA method has been proven to achieve a better trade-off between communication and sensing. The minimum CE rate was maximized while satisfying the localization accuracy constraints. To achieve this, the robust dual-functional precoders were designed using the \emph{semidefinite relaxation} (SDR), the sequential rank-$1$ constraint relaxation, and the \emph{successive convex approximation} (SCA) methods. Wang \etal \cite{10851844} extended the LEO ISAC system to more complex multi-satellite collaboration scenarios. By collecting the echo signals reflected by the target at all LEO satellites, a particle swarm optimization-based direct localization algorithm was proposed. To further improve the sensing performance while maintaining the communication performance, a joint beamforming problem was formulated. The transmit beamforming vectors were solved by using the SDR technique, the difference of convex algorithm, and the SCA method.

Few works have introduced MA into LEO systems for increasing the freedom of system design, thereby improving the system performance. For instance, Zhu \etal \cite{10806489} utilized the MA arrays to enhance the satellite beam coverage and interference mitigation. Specifically, the MA position vector and antenna weight vector were jointly optimized to minimize the average signal leakage power to the interference area of the satellite. Meanwhile, the average beamforming gain over the coverage area was designed to exceed the minimum threshold. The \emph{alternating optimization} (AO) algorithm and SCA technique were employed to solve the ISAC problem iteratively. A low-complexity substitute was also designed. However, to the best of our knowledge, no existing study has utilized MA to assist the LEO ISAC systems.

\subsection{Contributions}
In this paper, we investigate the improvement of the analytical localization accuracy in MA-assisted LEO ISAC systems while ensuring communication performance. An SPEB minimization problem is formulated and solved efficiently. The main contributions of this paper are outlined below:

\begin{itemize}
    \item We consider an MA-assisted LEO ISAC system where the LEO satellites move along a fixed orbit. We divide the LEOs' trajectory into multiple time slots, and enhance the sensing performance of multiple \emph{sensing equipments} (SEs) while ensuring communication performance within the communication coverage area in each time slot. We built the system geometric model. Several coordinate systems are established, and the transition matrices between them are derived.

    \item We derive the communication performance metric, the SINR, and the sensing performance metric, the SPEB, respectively. Based on these metrics, an SPEB minimization problem with SINR constraints, total transmit power constraint, and several MA physical constraints is formulated. Using the SDR, we make the original problem more tractable and then divide it into two subproblems that can be solved efficiently in an alternating manner. We also analyze the convergence and computational complexity of the proposed algorithm.

    \item We perform comprehensive numerical simulations to assess the performance of the proposed algorithm, and also benchmark the proposed algorithm against the state-of-the-art method in different dimensions. It is shown that the proposed algorithm can enhance communication performance across most of the communication coverage area. Meanwhile, it consistently improves the sensing performance by at least $25 \%$ compared to the benchmarks.
\end{itemize}

The rest of this paper is structured as follows. Section~\ref{section: SystemModel} presents the system model, including the geometric model, the transmit signal model, and the received signals at communication and sensing receivers, respectively. In Section~\ref{section:metric}, the communication and sensing performance metrics are derived, and the considered problem is formulated. The problem is efficiently solved in Section~\ref{section:Proposed}. Section~\ref{section:Results} showcases simulation results. The paper is concluded in Section~\ref{section:conclusion}.

\emph{Notation}: ${\rm Tr} \bigBraces{\cdot}$, $(\cdot)^\top$, $(\cdot)^*$, $(\cdot)^{\rm H}$, and $(\cdot)^{\rm -1}$ denote trace, transpose, conjugate, hermitian, and inverse, respectively. $\Vert \cdot \Vert$ denotes the $2$-norm. $\rank(\mathbf{A})$ denotes the rank of matrix $\mathbf{A}$. $\mathbf{A} \succcurlyeq \bm{0}$ indicates a positive semi-definite matrix $\mathbf{A}$. $\mathbf{I}_M$, $\mathbf{O}_{M \times N}$, and $\bm{0}_M$ denote $M \times M$ identity matrices, $M \times N$ all-zero matrices, and $M \times 1$ all-zero vectors, respectively. $\Re \bigBraces{\cdot}$ and $\Im \bigBraces{\cdot}$ represent the real and imaginary components, respectively. $\mathbb{E} (\cdot)$ denotes the expectation. $\mathcal{CN}(\mu, \sigma^2)$ represents the complex Gaussian distribution with a mean $\mu$ and variance $\sigma^2$. The notation used in this paper is listed in Table \ref{tab_notation}.

\begin{table}
\caption{Notation and definition.}
\label{tab_notation}
\centering
\begin{tabular}{| c | l |}
\hline
Notation & Definition \\
\hline
$\svTrans$ &  Steering vector of the MA array at LEO satellite \\
$\coverageArea$ & Communication coverage area\\
$\svRec$ &  Steering vector of the receiving array at SEs \\
$\mathcal{C}$ & Movable region of the MAs \\
$\dmin$ & Minimum spacing constraint of the MAs \\
$G_\text{e}$ & The constant of gravitation \\
$G_{\rm LEO}$ & Antenna gain at the LEO satellite \\
$G_{\rm CE}$ & Antenna gain at the CEs \\
$H_\text{s}$ & Altitude of the LEO \\
$J$ & Number of transmit signal samples \\
$\mathbf{J}_{\xiMK}$ & $\xiMK$-related FIM \\
$\mathbf{J}_{\bm{\eta}_{m,k}}$ & $\bm{\eta}_{m,k}$-related FIM \\
$K$ & Number of SEs \\
$K_\text{s}$ & Number of LEOs in one orbital plane \\
$L$ & Number of grids within $\coverageArea$ \\
$M$ & Number of slots \\
$M_\text{e}$ & Mass of the earth \\
$\Nt$ & Number of transmit antennas at LEO satellite \\
$\Nr$ & Number of receiving antennas at SEs \\
$\pmax$ & Total transmit power constraint \\
$\pLEO$ & Position of LEO satellite at the $m$-th slot \\
$\pSEk$ & Position of the $k$-th SE \\
$\pCEl$ & Position of the center of the $l$-th grid \\
$\qm$ & Position vector of the MAs \\
$R_\text{e}$ & Radius of the earth \\
$\Rsm$ & Covariance matrix of sensing signal at the $m$-th slot \\
${\rm SPEB}_{m,k}$ & SPEB of the $k$-th SE at the $m$-th slot \\
$T$ & Considered typical time interval \\
$T_\text{s}$ & Orbital period \\
$\Ta$ & Transition matrix between SCCS and GCCS \\
$\Tb$ & Transition matrix between GCCS and NEU \\
$\vmax$ & Moving speed constraint for MAs \\
$\wcm$ & Communication beamforming vector at the $m$-th slot \\
$\alphaMK$ & Complex amplitude at the $k$-th SE at the $m$-th slot  \\
$\beta$ & Orbital inclination angle \\
$\gamma$ & Path loss exponent \\
$\Gammal$ & SINR constraint for the $l$-th imaginary CE \\
$\bm{\eta}_{m,k}$ & Unknown parameter vector under GCCS \\
$\lambda$ & Signal wavelength \\
$\xiMK$ & Unknown parameter vector under GSCS \\
$\Ximk$ & Transition matrix between FIMs \\
$\senNoisePower$ & Noise power at SEs \\
$\commNoisePower$ & Noise power at imaginary CEs \\
\hline
\end{tabular}
\end{table}

\section{System Model}
\label{section: SystemModel}
In this section, we introduce the system model of the considered MA-assisted LEO ISAC system. Specifically, we first propose the geometric model, including the definition of several coordinate systems and the derivation of the transition matrices between them. Then, we define the joint communication and sensing transmit signal and present the received signals at the communication and sensing receivers, respectively. The received signals are used to derive communication and sensing metrics in Sec. \ref{section:metric}, respectively.

\begin{figure}[!t]
\centering
\subfloat[Illustration of the system geometric model and GCCS.]{\includegraphics[width=0.9\linewidth]{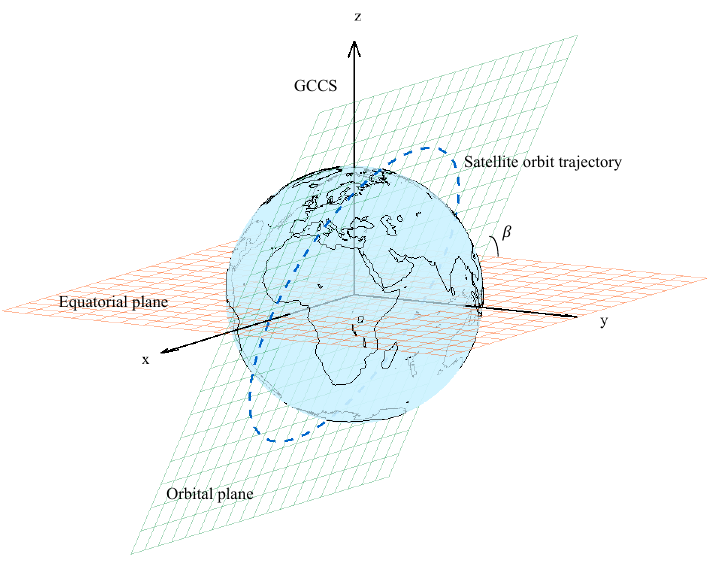}\label{FigSystemModel1}}
\hfil
\subfloat[Illustration of the SCCS and NEU coordinate system at SEs.]{\includegraphics[width=0.9\linewidth]{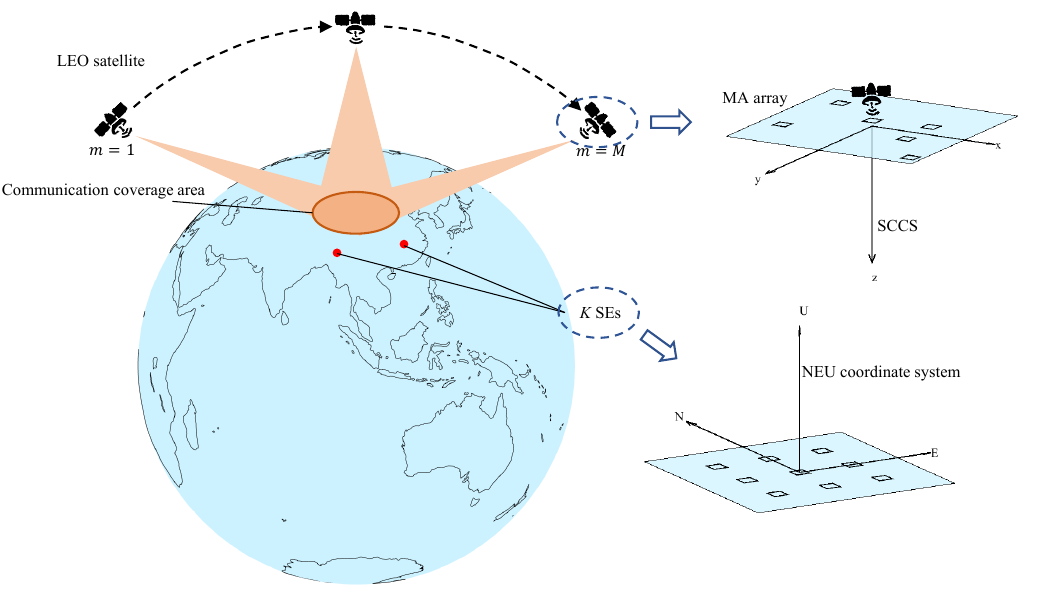}\label{FigSystemModel2}}
\caption{Illustration of the considered MA-assisted LEO ISAC system.}
\label{FigSystemModel}
\end{figure}

\subsection{Geometric Model}
As illustrated in Fig. \ref{FigSystemModel}(a), $K_\text{s}$ equidistantly distributed LEO satellites are considered in each orbital plane. The orbital inclination angle is $\beta$. \emph{Without loss of generality} (w.l.o.g.), we suppose the earth is a regular sphere with radius $R_\text{e}$. The altitude of the LEO satellites with respect to the earth's surface is $H_\text{s}$. Thus, the orbital period $T_\text{s} $ can be calculated by $T_\text{s} = 2 \pi \sqrt{\frac{(R_\text{e} + H_\text{s})^3}{G_\text{e} M_\text{e}}}$, where $G_\text{e}$ and $M_\text{e}$ are the constant of gravitation and the mass of the earth, respectively.

Since the $K_\text{s}$ LEO satellites in one orbital plane share the same trajectory, we focus on their motion with one typical time interval $(0, T]$ w.l.o.g., where $T = T_\text{s}/K_\text{s}$. Furthermore, we divide this interval into $M$ time slots. The number of $M$ is chosen to be sufficiently large such that the LEO satellite can be considered stationary with respect to an observer on the ground. Then, we use the snapshot at time $t=\frac{(m-1/2)T}{M},m=1,\ldots,M$, to represent the average performance of the $m$-th time slot. The position of the LEO satellite under the \emph{geocentric Cartesian coordinate system} (GCCS) at the $m$-th slot is $\pLEO = \left[ x_{{\rm LEO},m}, y_{{\rm LEO},m}, z_{{\rm LEO},m} \right]^\top$, where 
\begin{subequations} 
    \begin{align}
        x_{{\rm LEO},m} &= R_m \cos \Theta_m \cos \Phi_m; \\
        y_{{\rm LEO},m} &= R_m \cos \Theta_m \sin \Phi_m; \\
        z_{{\rm LEO},m} &= R_m \sin \Theta_m,
    \end{align}
\end{subequations}
with
\begin{subequations} 
    \begin{align}
        R_m &= R_\text{e} + H_\text{s}; \\
        \Theta_m &= \arcsin \braces{\sin \beta \sin \alpha_m}; \\
        \Phi_m &= \arctan \braces{\cos \beta \tan \alpha_m},
    \end{align}
\end{subequations}
and $\alpha_m$ being the angle traversed by the LEO satellite from the initial moment to the $m$-th time slot. Note that $\midBraces{R_m, \Theta_m, \Phi_m}^\top$ denotes the position of the LEO satellite under the \emph{geocentric spherical coordinate system} (GSCS), where $\Theta_m \in [-\pi/2, \pi/2]$, and $\Phi_m \in (-\pi, \pi]$.

A planar array composed of $\Nt$ MAs is deployed on the LEO satellite. We represent the MA positions under the \emph{satellite-centric Cartesian coordinate system} (SCCS) at the $m$-th slot as $\qm$. Note that the $x$ axis is towards the moving direction of the satellite, and the $z$ axis is towards the geocenter in SCCS. The direction of the $y$ axis can be determined to form a left-handed cartesian coordinate, as illustated in Fig. \ref{FigSystemModel}(b). The vector $\qm \in \mathbb{C}^{2 \Nt \times 1}$ concatenates the positions of all MAs in order, i.e., $\qm \triangleq \left[ \bm{q}_{m,1}^\top, \ldots, \bm{q}_{m, \Nt}^\top \right]^\top$. The position for the $n$-th MA at the $m$-th slot can be further expressed as $\bm{q}_{m,n} = \left[ \xnm, \ynm \right]^\top$. 

For consistent coordinate representation, the transition matrix from the SCCS to the GCCS at the $m$-th slot, $\Ta$, is given in (\ref{Ta}), at the top of the next page. Given the position of the LEO satellite, $\pLEO$, and a point in the 3D space, whose position is expressed by $\bm{p}_{\rm SCCS}$ under the SCCS, and $\bm{p}_{\rm GCCS}$ under the GCCS, $\bm{p}_{\rm SCCS}$ and $\bm{p}_{\rm GCCS}$ are interconvertible by
\begin{subequations} 
    \begin{align}
        \bm{p}_{\rm GCCS} &= \Ta \bm{p}_{\rm SCCS} + \pLEO; \\
        \bm{p}_{\rm SCCS} &= \Ta^\top \braces{\bm{p}_{\rm GCCS} - \pLEO}.
    \end{align}
\end{subequations}

\begin{figure*}[!t]
\normalsize
\begin{equation}
    \label{Ta}
    \Ta = 
    \begin{bmatrix}
		- \sin \beta \sin \Theta_m - \cos \beta \cos \Theta_m \sin \Phi_m & 0 & - \cos \Theta_m \cos \Phi_m \\
		\cos \beta \cos \Theta_m \cos \Phi_m & \sin \beta & -\cos \Theta_m \sin \Phi_m \\
		\sin \beta \cos \Theta_m \cos \Phi_m & - \cos \beta & -\sin \Theta_m
    \end{bmatrix}
\end{equation}
\begin{equation}
    \label{Tb}
	\Tb =
    \begin{bmatrix}
		- \sin \Phi_r & - \cos \Phi_r \sin \Theta_r & \cos \Phi_r \cos \Theta_r \\
		\cos \Phi_r & - \sin \Phi_r \sin \Theta_r &  \sin \Phi_r \cos \Theta_r \\
		0 & \cos \Theta_r & \sin \Theta_r
	\end{bmatrix}
\end{equation}
\hrulefill
\vspace*{-5pt}
\end{figure*}

As illustrated in Fig. \ref{FigSystemModel}(b), in the considered ISAC scenario, suppose there are $K$ \emph{sensing equipments} (SEs) on the earth's surface. The receiver on each SE is a \emph{uniform planar array} (UPA)  composed of $\Nr$ antennas. At each SE, we construct an \emph{East-North-Up} (NEU) coordinate system, as illustrated in Fig. \ref{FigSystemModel}(b). Given the position of the origin of one NEU, $\bm{p}_{\text{r},k}, k=1,\ldots,K$, under GCCS, and a point in the 3D space whose position is $\bm{p}_{\rm GCCS}$ under the GCCS, and $\bm{p}_{\rm NEU}$ under the NEU, $\bm{p}_{\rm GCCS}$ and $\bm{p}_{\rm NEU}$ are also interconvertible by
\begin{subequations}
    \begin{align}
        \bm{p}_{\rm GCCS} &= \Tb \bm{p}_{\rm NEU} + \bm{p}_{\rm r}; \\
        \bm{p}_{\rm NEU} &= \Tb^\top \braces{\bm{p}_{\rm GCCS} - \bm{p}_{\rm r}},
    \end{align}
\end{subequations}
where the transition matrix $\Tb$ is given in (\ref{Tb}), at the top of the next page.

On the other hand, we aim to maintain the communication performance within a common coverage area $\coverageArea$ during the $M$ slots. To achieve this, we equally divide the longitude and the latitude of the earth into $L_\text{a} \times L_\text{e}$ grids. The performance at grid centers is used to represent the overall performance within grids. Suppose there are $L$ grids within $\coverageArea$, and the center position of the $l$-th grid is $\pCEl, l=1,\ldots,L$. We use the performance at the $L$ grid centers to represent the overall performance in $\coverageArea$.

\subsection{Transmit Signal Model}

To carry out sensing and communication simultaneously, the MA array at the LEO satellite transmit the joint signal at the time index $\timeIndex$ in the $m$-th slot as
\begin{equation}
    \bm{x}(\timeIndex) = \wcm c(\timeIndex) + \bm{s}(\timeIndex),
\end{equation}
where $\wcm \in \mathbb{C}^{\Nt \times 1}$ denotes the transmit beamforming vector for broadcasting the information stream $c(\timeIndex)$ to the communication coverage area $\coverageArea$. The signal $\bm{s}(\timeIndex)$ is the dedicated sensing signal with covariance matrix $\Rsm = \mathbb{E} \left[ \bm{s}(\timeIndex) \bm{s}^{\rm H}(\timeIndex)\right]$. Furthermore, the communication signals are independent Gaussian random signals with zero mean and unit power, and $\bm{s}(\timeIndex)$ is generated by pseudo-random coding. Thus, $\mathbb{E} \left[ c(\timeIndex) c^*(\timeIndex)  \right] = 1$ and $\mathbb{E} \left[ c(\timeIndex) \bm{s}(\timeIndex)^{\rm H} \right] = \bm{0}_{\Nt}^\top$. The covariance matrix of $\bm{x}(\timeIndex)$ can be given by
\begin{equation}
    \mathbf{R}_{x,m} = \mathbb{E}\left[ \bm{x}(\timeIndex) \bm{x}^{\rm H}(\timeIndex) \right] = \wcm \wcm^{\rm H} + \Rsm.
\end{equation}
In practice, $\mathbf{R}_{x,m}$ is calculated by averaging over $J$ time indexes
\begin{equation}
    \label{Rx}
    \Rxm \approx \frac{1}{J} \mathbf{X} \mathbf{X}^{\rm H},
\end{equation}
where $\mathbf{X} = \left[ \bm{x}(1), \ldots, \bm{x}(J) \right]$. In this paper, we adopt a sufficiently large $J$ so that the approximation in (\ref{Rx}) is accurate.

\subsection{Received Communication Signal Model}

Suppose there are imaginary \emph{communication equipments} (CEs) located at the $L$ grid centers within $\coverageArea$. The received signal at the $l$-th CE at the $m$-th slot can be modeled as
\begin{equation}
    \label{yc}
    \yc(\timeIndex) = \sqrt{\rhoml} \aml^\top \bm{x}(\timeIndex) + n_l,
\end{equation}
where $\rhoml = G_{\rm LEO} G_{\rm CE} \braces{\frac{4 \pi}{\lambda}}^2 d_{m,l}^{-\gamma}$ is the path loss between the LEO satellite and the center of the $l$-th grid at the $m$-th slot; $G_{\rm LEO}$ and $G_{\rm CE}$ are the antenna gain at the LEO satellite and the $l$-th CE, respectively; $\aml \triangleq \svTrans(\pCEl)$; $\lambda$ is the system wavelength; $d_{m,l}$ is the distance between the LEO satellite and the $l$-th CE at the $m$-th slot; $\gamma$ is the path loss exponent. Given a position $\bm{p}$ under GCCS, the steering vector of the MA array at the $m$-th slot towards this position's direction is denoted by $\svTrans(\bm{p})$. The $n$-th element of $\svTrans(\bm{p})$ can be expressed by
\begin{equation}
    \svTrans(\bm{p})[n] = \exp \bigBraces{j \frac{2 \pi}{\lambda} \frac{\braces{\bm{p} - \pLEO}^\top}{\VertSquare{\bm{p} - \pLEO}} \Ta \qnm},
\end{equation}
where $n=1,\ldots,\Nt$. $n_l$ represents the complex \emph{additive white Gaussian noise} (AWGN) at the $l$-th CE with zero mean and variance $\commNoisePower$.

\subsection{Received Sensing Signal Model}

Suppose there are $K$ \emph{sensing equipments} (SEs) located at $\pSEk, k=1,\ldots,K$, respectively. 
The received signal at the $k$-th SE at the $m$-th slot can be modeled as
\begin{align}
    \label{Ys}
    \Ys = \alphaMK \bmk \amk^\top \X + \Ns,
\end{align}
where $\alphaMK$ stands for the complex amplitude of the received signal, and is considered deterministic unknown \cite{1703855}; $\amk \triangleq \svTrans(\pSEk)$, and $\bmk \triangleq 
\svRec(\pLEO)$; matrices $\Ys \in \mathbb{C}^{\Nr \times J}$ and $\Ns \in \mathbb{C}^{\Nr \times J}$ horizontally concatenate the received sensing signals and AWGN over $J$ time indexes, respectively. Each entry within $\Ns$ obeys $\mathcal{CN}(0, \senNoisePower)$. Denote the position of the $\tilde{n}$-th antenna at the $k$-th SE under the SE's local NEU coordinate system as $\tilde{\bm{q}}_{\tilde{n},k}$. Given a position $\bm{p}$ under GCCS, the steering vector of UPA at the $k$-th SE towards the position's direction is denoted by $\svRec(\bm{p})$. The $\tilde{n}$-th element of $\svRec(\bm{p})$ can be expressed by
\begin{equation}
    \svRec(\bm{p})[\tilde{n}] = \exp \bigBraces{j \frac{2 \pi}{\lambda} \frac{(\bm{p} - \pSEk)^\top}{\Vert \bm{p} - \pSEk \Vert} \mathbf{T}_b \tilde{\bm{q}}_{\tilde{n},k}},
\end{equation}
where $\tilde{n} = 1, \ldots, \Nr$.

Furthermore, for the sake of deriving the sensing performance metric in Sec. \ref{subsec:SPEB}, the expression in (\ref{Ys}) is vectorized as 
\begin{equation}
    \ys = \us + \ns,
\end{equation}
where
$\us = {\rm vec} \braces{\alphaMK \svRec(\pLEO) \svTrans^\top(\pSEk) \X}$ and $\ns = {\rm vec} \braces{\Ns}$.

\section{Performance Metrics and Problem Formulation}
\label{section:metric}
In this section, we first derive the communication performance metric, i.e., \emph{signal-to-interference-plus-noise ratio} (SINR); and the sensing performance metric, i.e., SPEB, for the considered LEO ISAC system. Then, utilizing these metrics, we formulate the joint beamforming and MA position problem.
\subsection{Metric for Communication: Communication SINR}
We employ the commonly used SINR as a metric to assess the ISAC system's communication performance. To derive the SINR, we express the received signal at the $l$-th CE at the $m$-th slot in (\ref{yc}) as
\begin{align}
    \yc(\timeIndex) =& \underbrace{\sqrt{\rhoml} \aml^\top \wcm c(\timeIndex)}_\text{Desired Signal} \notag \\ 
    &+ \underbrace{\sqrt{\rhoml} \aml^\top \bm{s}(\timeIndex)}_\text{Sensing Interference} + n_l.
\end{align}
Accordingly, the SINR of the $l$-th CE, $\gamma_{m,l}$, at the $m$-th slot is expressed as
\begin{equation}
    \gamma_{m,l} = \frac{\rhoml |\aml^\top \wcm|^2 }{\rhoml \aml^\top \Rsm \aml^* + \commNoisePower }.
\end{equation}

\subsection{Metric for Sensing: SPEB}
\label{subsec:SPEB}
To evaluate the sensing performance of the $k$-th SE, we derive the SPEB for estimating $\pSEk$, which provides a lower bound for the \emph{mean square error} (MSE) and has a closed-form expression. Since the SE is located on the earth's surface and has a fixed range, $R_\text{e}$, to the origin of the GSCS, its position can be determined by the parameters $\ThetaSEk$ and $\PhiSEk$, which are the elevation and azimuth angles of the SE under the GSCS, respectively.

We define the unknown parameter vector $\xiMK \triangleq \midBraces{\bm{\phi}_k^\top, \bmAlphaMK^\top}^\top$ at the $k$-th SE at the $m$-th slot, where $\bm{\phi}_k \triangleq \midBraces{\ThetaSEk, \PhiSEk}^\top$, and $\bm{\alpha}_{m,k} \triangleq \midBraces{\Re\bigBraces{\alphaMK}, \Im \bigBraces{\alphaMK}}^\top$. Then, the FIM \cite{kay1993fundamentals} for estimating $\xiMK$ can be partitioned as
\begin{equation}
    \mathbf{J}_{\xiMK} =
    \begin{bmatrix}
		\mathbf{J}_{\phiK \phiK} & \mathbf{J}_{\phiK \bmAlphaMK} \\
		\mathbf{J}_{\phiK \bmAlphaMK}^\top & \mathbf{J}_{\bmAlphaMK \bmAlphaMK}
	\end{bmatrix}.
\end{equation}
$\mathbf{J}_{\phiK \phiK}$ can be further partitioned as
\begin{equation}
    \mathbf{J}_{\phiK \phiK} =
    \begin{bmatrix}
        J_{\ThetaSEk \ThetaSEk} & J_{\ThetaSEk \PhiSEk} \\
        J_{\ThetaSEk \PhiSEk}^\top & J_{\PhiSEk \PhiSEk}
    \end{bmatrix}.
\end{equation}
The entries $J_{pq}, \forall p,q \in \bigBraces{\ThetaSEk, \PhiSEk}$, $\mathbf{J}_{\phiK \bmAlphaMK}$ and $\mathbf{J}_{\bmAlphaMK \bmAlphaMK}$ can be calculated as in (\ref{Jpq}), (\ref{JPhiAlpha}), and (\ref{JAlphaAlpha}), respectively, at the top of this page.
\begin{figure*}[!t]
\normalsize
\begin{align}
    \label{Jpq}
    J_{pq} &= \frac{2}{\senNoisePower} \Re \bigBraces{ \alphaMK^* {\rm vec} \braces{\braces{\partials{\bmk}{p} \amk + \bmk \partials{\amk^\top}{p}} \X}^{\rm H} \alphaMK {\rm vec} \braces{\braces{ \partials{\bmk}{q} \amk^\top + \bmk \partials{\amk^\top}{q} \X} } } \notag	\\ 
    &=\frac{2 |\alphaMK|^2 J}{\senNoisePower} \Re \bigBraces{ {\rm Tr} \bigBraces{\braces{ \partials{\bmk}{q} \amk^\top + \bmk \partials{\amk^\top}{q}} \Rxm \braces{\partials{\amk^*}{p} \bmk^{\rm H} + \amk^* \partials{\bmk^{\rm H}}{p}} }};
\end{align}
\begin{align}
    \label{JPhiAlpha}
    \mathbf{J}_{\phiK\bmAlphaMK} &= \frac{2}{\senNoisePower} \Re \bigBraces{ 
    \begin{bmatrix}
        \partials{\us^{\rm H}}{\ThetaSEk} \\
        \partials{\us^{\rm H}}{\PhiSEk}
    \end{bmatrix}
    {\rm vec} \braces{\bmk \amk^\top \X} \midBraces{1, j}
    } \notag \\
    &= \frac{2}{\senNoisePower} \Re \bigBraces{
    \begin{bmatrix}
        \alphaMK^* {\rm vec} \braces{\braces{\partials{\bmk}{\ThetaSEk} \amk^\top + \bmk \partials{\amk^\top}{\ThetaSEk} } \X}^{\rm H} \\
        \alphaMK^* {\rm vec} \braces{\braces{\partials{\bmk}{\PhiSEk} \amk^\top + \bmk \partials{\amk^\top}{\PhiSEk} } \X}^{\rm H} 
    \end{bmatrix}
    {\rm vec} \braces{\bmk \amk^\top \X} \midBraces{1, j} 
    } \notag \\
    &= \frac{2 J}{\senNoisePower} \Re \bigBraces{
    \begin{bmatrix}
        \alphaMK^* {\rm Tr} \bigBraces{\bmk \amk^\top \Rxm \braces{\partials{\amk^*}{\ThetaSEk} \bmk^{\rm H} + \amk^* \partials{\bmk^{\rm H}}{\ThetaSEk}}} \\
        \alphaMK^* {\rm Tr} \bigBraces{\bmk \amk^\top \Rxm \braces{\partials{\amk^*}{\PhiSEk} \bmk^{\rm H} + \amk^* \partials{\bmk^{\rm H}}{\PhiSEk}}} 
    \end{bmatrix} \midBraces{1, j}
    };
\end{align}
\begin{align}
    \label{JAlphaAlpha}
    \mathbf{J}_{\bmAlphaMK \bmAlphaMK} &= \frac{2}{\senNoisePower} \Re \bigBraces{\partials{\us^{\rm H}}{\bmAlphaMK} \partials{\us}{\bmAlphaMK}} \notag \\ 
    &= \frac{2}{\senNoisePower} \Re \bigBraces{ \braces{{\rm vec}\braces{\bmk \amk^\top \X} \midBraces{ 1, j}}^{\rm H} {\rm vec} \braces{\bmk \amk^\top \X} \midBraces{1, j} } \notag \\ 
    &= \frac{2}{\senNoisePower} \mathbf{I}_2 {\rm Tr} \bigBraces{\bmk \amk^\top \X \X^{\rm H} \amk^* \bmk^{\rm H}} \notag \\ 
    &= \frac{2 J}{\senNoisePower} \mathbf{I}_2 {\rm Tr} \bigBraces{\bmk \amk^\top \Rxm \amk^* \bmk^{\rm H}}.
\end{align}
\hrulefill
\vspace*{-5pt}
\end{figure*}

To calculate the SPEB for the sensing problem, we define a new parameter vector $\bm{\eta}_{m,k} = \midBraces{\pSEk^\top, \bmAlphaMK^\top}^\top$, which is related to the GCCS position of the $k$-th SE. The $\xiMK$-related FIM, i.e., $\mathbf{J}_{\xiMK}$, can be transformed into a $\bm{\eta}_{m,k}$-related FIM, $\mathbf{J}_{\bm{\eta}_{m,k}}$, by
\begin{equation}
    \mathbf{J}_{\bm{\eta}_{m,k}}^{-1} = \Ximk \mathbf{J}_{\xiMK}^{-1} \Ximk^\top,
\end{equation}
where $\Ximk= \mathbf{E} \mathbf{T}_{m,k}$ with $\mathbf{E} = \begin{bmatrix}
		\mathbf{I}_3 & \mathbf{O}_{2 \times 2}
\end{bmatrix}$ being the extraction matrix and $\mathbf{T} = \partials{\bm{\eta}_{m,k}}{\xiMK}$ being the transition matrix. Then, the SPEB is given by
\begin{equation}
    \SPEB = {\rm Tr} \bigBraces{ \mathbf{J}_{\bm{\eta}_{m,k}}^{-1}},
\end{equation}
and the PEB can be directly calculated by ${\rm PEB}_{m,k}=\sqrt{\SPEB}$.

\subsection{Problem Formulation}
\label{ProblemFormulation}
Based on the derived communication SINR and sensing SPEB, our objective is to improve the average sensing performance as much as possible while maintaining the minimum SINR requirement in each slot. The overall problem at the $m$-th slot is formulated as
\begin{subequations}
\label{R1}
\begin{align}
    \label{R1a}
    (\rm P1): \quad &\underset{ \qm, \wcm, \Rsm }{\text{min}} \quad \sum\limits_{k=1}^K \SPEB  \\
    \label{R1b}
    \text{s.t.} \quad &  {\rm Tr} \left\{\wcm \wcm^{\rm H} + \Rsm \right\} \leq \pmax, \\
    \textbf{\label{R1c}}
    & \gamma_l \geq \Gamma_l , \forall l, \\
    \label{R1d}
    & \qnm \in \mathcal{C}, \forall n, \\
    \label{R1e}
    & \Vert \qnm - \qnhatm \Vert \geq \dmin, \forall n \neq \hat{n}, \\
    \label{R1f}
    & \Vert \qnm - \bm{q}_{m-1,n} \Vert \leq \frac{\vmax T}{M}, \\
    \label{R1g}
    & \Rsm \succcurlyeq \bm{0},
\end{align}
\end{subequations}
where (\ref{R1b}) limits the total transmit power; (\ref{R1c}) is the communication SINR constraints for all CEs; (\ref{R1d}) is the moving region constraint for MAs; (\ref{R1e}) guarantees a minimum inter-antenna spacing; (\ref{R1f}) restrains the maximum moving speed of each MA; (\ref{R1g}) is the semidefinite constraint of the covariance matrix of the sensing signal. We propose a novel alternative approach to solve the highly non-convex problem (P1) in the next section.

\section{Joint beamforming and MA position design}
\label{section:Proposed}
In this section, we jointly optimize the transmit beamforming and MA positions. Specifically, we reformulate the original problem (P1) into a more tractable form based on the \emph{semidefinite relaxation} (SDR). Subsequently, we present an efficient AO algorithm that alternately updates the transmit beamforming vector and MA positions at each time slot. At last, the convergence and complexity are analyzed.
\subsection{Problem Reformulation}
To make the original problem (P1) more tractable, we define $\Wcm \triangleq \wcm \wcm^{\rm H}$ with $\Wcm \succcurlyeq 0$ and $\rank(\Wcm) = 1$. Now, we aim to transform the problem (P1) into an equivalent problem with respect to $\qm$, $\Wcm$, and $\Rsm$.

We start with the objective function (\ref{R1a}). According to the expression of the entries of $\mathbf{J}_{\xiMK}$ in (\ref{Jpq}), (\ref{JPhiAlpha}), and (\ref{JAlphaAlpha}), $\mathbf{J}_{\xiMK}$ is linear to $\Rxm$, which satisfies $\Rxm = \Wcm + \Rsm$. Thus, (\ref{R1a}) can be expressed by $\qm$, $\Wcm$, and $\Rsm$. Next, (\ref{R1b}) is tranformed into ${\rm Tr} \bigBraces{\Wcm + \Rsm} \leq \pmax$. The constraint (\ref{R1c}) can be expanded into: $l=1,\ldots,L$,
\begin{equation}
    \frac{\rhoml \aml^\top \Wcm \aml^* }{\rhoml \aml^\top \Rsm \aml^* + \commNoisePower } \geq \Gammal.
\end{equation}
So far, the objective function and all constraints in (P1) have been reconstructed with respect to $\qm$, $\Wcm$, and $\Rsm$. Thus, a new equivalent problem of (P1) is
\begin{subequations}
\label{R2}
\begin{align}
    \label{R2a}
    (\rm P2): \quad &\underset{ \qm, \Wcm, \Rsm }{\text{min}} \quad \sum\limits_{k=1}^K \SPEB \\
    \label{R2b}
    \text{s.t.} \quad & {\rm Tr} \left\{\Wcm + \Rsm \right\} \leq \pmax, \\
    \label{R2c}
    & \frac{\rhoml \aml^\top \Wcm \aml^* }{\rhoml \aml^\top \Rsm \aml^* + \commNoisePower } \geq \Gammal, \forall l, \\
    \label{R2d}
    & \Wcm \succcurlyeq \bm{0}, \\
    & \Rsm \succcurlyeq \bm{0}, \\
    \label{R2e}
    & \rank(\Wcm) = 1,\\
    & \qnm \in \mathcal{C}, \forall n, \\
    & \Vert \qnm - \qnhatm \Vert \geq \dmin, \forall n \neq \hat{n}, \\
    & \Vert \qnm - \bm{q}_{m-1,n} \Vert \leq \frac{\vmax T}{M}.
\end{align}
\end{subequations}

To tackle the non-convex problem (P2) efficiently, we adopt the SDR method to relax the non-convex rank-$1$ constraint (\ref{R2e}). Then, based on AO, we decompose (P2) into two sub-problems, solved in Sec. \ref{subsection:TBF}, and Sec. \ref{subsection:pMA}, respectively.

\subsection{Optimization of Transmit Beamforming}
\label{subsection:TBF}
Given fixed $\qm$, $\Wcm$ and $\Rsm$ can be solved by
\begin{subequations}
\label{Sub1Origin}
\begin{align}
(\rm P3): \quad & \underset{\Wcm, \Rsm}{\text{min}} \quad \sum\limits_{k=1}^K {\rm SPEB}_{m,k} \\
\text{s.t.} \quad & \text{(\ref{R2b}), (\ref{R2c}), (\ref{R2d}).}\notag
\end{align}
\end{subequations}
By defining $\Aml \triangleq \aml^* \aml^\top$, we can further convert (\ref{R2c}) into
\begin{equation}
    \frac{1}{\Gammal}{\rm Tr} \bigBraces{\Aml \Wcm} \geq {\rm Tr} \bigBraces{\Aml \Rsm} + \frac{\commNoisePower}{\rhoml}, \forall l,
\end{equation}
which is a convex constraint. However, the problem (P3) is still difficult to solve due to the non-convex objective function. To tackle this, we define $K$ auxiliary variables $\mathbf{Z}_{m,1}, \ldots \mathbf{Z}_{m,K}$ that satisfy
\begin{equation}
    \label{Zk}
    \Zmk \succcurlyeq \Ximk \mathbf{J}_{\xiMK}^{-1} \Ximk^\top, k=1,\ldots,K,
\end{equation}
which would result in ${\rm Tr} \bigBraces{\Zmk} \geq {\rm SPEB}_{m,k}$. Then, according to the property of the Schur Complement, (\ref{Zk}) can be transformed into
\begin{equation}
    \begin{bmatrix}
        \mathbf{Z}_{m,k} & \Ximk \\
        \Ximk^\top & \mathbf{J}_{\bm{\xi}_k}
    \end{bmatrix} \succcurlyeq \bm{0}, \forall k.
\end{equation}

Therefore, problem (P3) is equivalent to
\begin{subequations}
    \label{Sub1Final}
    \begin{align}
		(\rm P4): \quad & \underset{\Wcm, \Rsm, \bigBraces{\Zmk}_{\forall k}}{\text{min}} \quad  \sum\limits_{k=1}^K {\rm Tr} \bigBraces{\Zmk}   \\
		\text{s.t.} \quad & {\rm Tr} \left\{\Wcm + \Rsm \right\} \leq \pmax, \\
		& \frac{1}{\Gammal}{\rm Tr} \bigBraces{\Aml \Wcm} \geq {\rm Tr} \bigBraces{\Aml \Rsm} + \frac{\commNoisePower}{\rhoml}, \forall l, \\
		\label{LMI1}
        & \begin{bmatrix}
            \mathbf{Z}_{m,k} & \Ximk \\
            \Ximk^\top & \mathbf{J}_{\bm{\xi}_k}
        \end{bmatrix} \succcurlyeq \bm{0}, \forall k, \\
        \label{LMI2}
        & \Wcm \succcurlyeq \bm{0}, \\
        \label{LMI3}
        & \Rsm \succcurlyeq \bm{0}.
	\end{align}
\end{subequations}
Problem (P4) is a convex SDP problem, and can be solved efficiently, e.g., using CVX toolboxes~\cite{grant2014cvx}.

\subsection{Optimization of MA Positions}
\label{subsection:pMA}
Given fixed $\Wcm$ and $\Rsm$, $\qm$ can be solved by
\begin{subequations}
\label{Sub2Origin}
\begin{align}    
    (\rm P5): \quad & \underset{\qm}{\text{min}} \quad \sum\limits_{k=1}^K \SPEB \\
    \text{s.t.} \quad & \text{(\ref{R1c}), (\ref{R1d}), (\ref{R1e}), (\ref{R1f}).}\notag
\end{align}
\end{subequations}
Since the objective function of (P5) is highly non-convex with respect to $\qm$, we design a descent direction search method to iteratively approach the local optimal solution. Define function $f(\qm) \triangleq \sum\limits_{k=1}^K \SPEB$. During the $i$-th iteration, we approximate $f(\qm)$ by its first-order Taylor expansion near $\qm^{[i-1]}$ as
\begin{align}
    f(\qm) \approx & f(\qm^{[i-1]}) + \sum\limits_{n=1}^{\Nt} \left( \partials{f(\qm)}{\xnm}(\xnm - \xnm^{[i-1]}) \right. \notag \\
    & \left. + \partials{f(\qm)}{\ynm}(\ynm - \ynm^{[i-1]}) \right).
\end{align}

Then, we handle the minimum inter-antenna spacing constraint (\ref{R1e}). For any two antennas, i.e., $\forall n \neq \hat{n}$, we derive the lower bound of their spacing, $\Vert \qnm - \qnhatm \Vert$, via the first-order Taylor expansion near $\Vert \qnm^{[i-1]} - \qnhatm^{[i-1]} \Vert$ as
\begin{align}
    \Vert \qnm - \qnhatm \Vert \geq & \Vert \qnm^{[i-1]} - \qnhatm^{[i-1]} \Vert + \frac{\left(\qnm^{[i-1]} - \qnhatm^{[i-1]}\right)^\top}{\Vert \qnm^{[i-1]} - \qnhatm^{[i-1]} \Vert} \notag \\
    & \times \left( (\qnm - \qnhatm) - (\qnm^{[i-1]} - \qnhatm^{[i-1]}) \right) \notag \\ 
	=& \frac{(\qnm^{[i-1]} - \qnhatm^{[i-1]})^\top (\qnm - \qnhatm)}{\Vert \qnm^{[i-1]} - \qnhatm^{[i-1]} \Vert} \notag \\
    \triangleq & \hat{d}(\qnm, \qnhatm).
\end{align}
Thus, the constraint (\ref{R1e}) is converted to 
\begin{equation}
    \label{transformedDminConstraint}
    \hat{d}(\qnm, \qnhatm) \geq \dmin, \forall n \neq \hat{n}.
\end{equation}

Finally, we handle the SINR constraints (\ref{R1c}). Define $\Gml \triangleq \aml^\top \Wcm \aml^*$ and $\Iml \triangleq \aml^\top \Rsm \aml^*$, the constraint (\ref{R1c}) can be rewritten as
\begin{equation}
    \label{SINRConstraint}
    \Gml - \Gammal \Iml \geq \frac{\commNoisePower \Gammal}{\rhoml}, \forall l.
\end{equation}

We start by reformatting the expression $\Gml$. Since $\Wcm$ is Hermitian, suppose $\rank(\Wcm)=\RankC$, then $\Gml$ can be decomposed into
\begin{equation}
    \Gml = \aml^\top \Wcm \aml^* = \sum\limits_{r=1}^{\RankC} \lambdaMR \vert \uMR^{\rm H} \aml^* \vert^2,
\end{equation}
where $\lambdaMR$ and $\uMR$ are the $r$-th eigenvalue and eigenvector of $\Wcm$, respectively. Denote the $n$-th element of $\uMR$, i.e., $\uMR[n], n=1,\ldots,\Nt$, by its amplitude $\bmrn$ and phase $\phimrn$ as $\uMR[n] = \bmrn e^{j \phimrn}$. Then, $\Gml$ can be further expanded as 
\begin{align}
    \label{Gml}
    &\Gml = \sum\limits_{r=1}^{\RankC} \lambdaMR \Big| \sum\limits_{n=1}^{\Nt}  \bmrn e^{-j \phimrn} e^{- j \tilde{\bm{v}}^\top \qnm} \Big|^2 \notag \\
    =& \sum\limits_{r=1}^{\RankC} \lambdaMR \sum\limits_{n=1}^{\Nt}\sum\limits_{\hat{n}=1}^{\Nt} \bmrn \bmrnHat \cosbraces{\vml^\top \deltaQ + \deltaPhi},
\end{align}
where $\vml \triangleq \braces{\frac{2 \pi}{\lambda} \frac{\braces{\pCEl - \pLEO}^\top}{\Vert \pCEl - \pLEO \Vert} \Ta}^\top$, $\deltaQ \triangleq\qnm - \qnhatm$, and $\deltaPhi \triangleq \phimrn - \phimrnHat$. Then, we approximate the cosine term in (\ref{Gml}) by its second-order Taylor expansion at $\deltaQPrev \triangleq \qnm^{[i-1]} - \qnhatm^{[i-1]}$ as in (\ref{cosTermInGl}), at the top of this page. Since $\cosbraces{\vml^\top \deltaQPrev + \deltaPhi} \leq 1$ always holds, the cosine term is globally lower-bounded, i.e., (\ref{cosTermInGlLowerBound}), at the top of this page.
\begin{figure*}[!t]
\normalsize
\begin{align}
    \label{cosTermInGl}
    \cosbraces{\vml^\top \deltaQ + \deltaPhi} \approx & \cosbraces{\vml^\top \deltaQPrev + \deltaPhi} - \sinbraces{\vml^\top \deltaQPrev + \deltaPhi} \vml^\top \braces{\deltaQ - \deltaQPrev} \notag \\ 
    & - \frac{1}{2} \cosbraces{\vml^\top \deltaQPrev + \deltaPhi} \braces{\vml^\top \braces{\deltaQ - \deltaQPrev}}^2 \\
    \label{cosTermInGlLowerBound}
    \geq & \cosbraces{\vml^\top \deltaQPrev + \deltaPhi} - \sinbraces{\vml^\top \deltaQPrev + \deltaPhi} \vml^\top \braces{\deltaQ - \deltaQPrev} \notag \\
    & - \frac{1}{2} \braces{\vml^\top \braces{\deltaQ - \deltaQPrev}}^2 \triangleq \GlLowerBound
\end{align}
\begin{align}
    \label{cosTermInIl}
    \cosbraces{\vml^\top \deltaQ + \deltaPhiForI} \approx & \cosbraces{\vml^\top \deltaQPrev + \deltaPhiForI} - \sinbraces{\vml^\top \deltaQPrev + \deltaPhiForI} \vml^\top \braces{\deltaQ - \deltaQPrev} \notag \\ 
    & - \frac{1}{2} \cosbraces{\vml^\top \deltaQPrev + \deltaPhiForI} \braces{\vml^\top \braces{\deltaQ - \deltaQPrev}}^2 \\
    \label{cosTermInIlUpperBound}
    \leq & \cosbraces{\vml^\top \deltaQPrev + \deltaPhiForI} - \sinbraces{\vml^\top \deltaQPrev + \deltaPhiForI} \vml^\top \braces{\deltaQ - \deltaQPrev} \notag \\
    & + \frac{1}{2} \braces{\vml^\top \braces{\deltaQ - \deltaQPrev}}^2 \triangleq \IlUpperBound
\end{align}
\hrulefill
\vspace*{-5pt}
\end{figure*}
Thus, 
\begin{equation}
    \label{GlInequality}
    \Gml \geq \sum\limits_{r=1}^{\RankC} \lambdaMR \sum\limits_{n=1}^{\Nt}\sum\limits_{\hat{n}=1}^{\Nt} \GlLowerBound.
\end{equation}

Similarly, suppose $\rank \braces{\Rsm} = \RankS$, then $\Iml$ can be decomposed into
\begin{equation}
    \Iml = \aml^\top \Rsm \aml^* = \sum\limits_{\tilde{r}=1}^{\RankC} \lambdaMRforI \vert \uMRforI^{\rm H} \aml^* \vert^2,
\end{equation}
where $\lambdaMRforI$ and $\uMRforI$ are the $\tilde{r}$-th eigenvalue and eigenvector of $\Rsm$, respectively. Denote the $n$-th element of $\uMRforI$, i.e., $\uMRforI[n], n=1,\ldots,\Nt$, by its amplitude $\bmrnHatForI$ and phase $\phimrnHatForI$ as $\uMRforI[n] = \bmrnForI e^{j \phimrnForI}$, $\Iml$ can be further expanded as
\begin{align}
    \label{Iml}
    &\Iml = \sum\limits_{r=1}^{\RankS} \lambdaMRforI \Big| \sum\limits_{n=1}^{\Nt}  \bmrnForI e^{-j \phimrnForI} e^{- j \tilde{\bm{v}}^\top \qnm} \Big|^2 \\ \notag
    =& \sum\limits_{r=1}^{\RankS} \lambdaMRforI \sum\limits_{n=1}^{\Nt}\sum\limits_{\hat{n}=1}^{\Nt} \bmrnForI \bmrnHatForI \cosbraces{\vml^\top \deltaQ + \deltaPhiForI},
\end{align}
where $\deltaPhiForI \triangleq \phimrnForI - \phimrnHatForI$. We again approximate the cosine term by its second-order Taylor expansion at $\deltaQPrev \triangleq \qnm^{[i-1]} - \qnhatm^{[i-1]}$ as in (\ref{cosTermInIl}), at the top of this page. Since $\cosbraces{\vml^\top \deltaQPrev + \deltaPhiForI} \geq -1$ always holds, the cosine term is globally upper-bounded, i.e., (\ref{cosTermInIlUpperBound}), at the top of this page. Thus, 
\begin{equation}
    \label{IlInequality}
    \Iml \leq \sum\limits_{\tilde{r}=1}^{\RankS} \lambdaMRforI \sum\limits_{n=1}^{\Nt}\sum\limits_{\hat{n}=1}^{\Nt} \IlUpperBound.
\end{equation}

According to the inequalities (\ref{GlInequality}) and (\ref{IlInequality}), the constraint (\ref{SINRConstraint}) can be transformed into
\begin{equation}
\label{transformedSINRConstraint}
\sum\limits_{n=1}^{\Nt}\sum\limits_{\hat{n}=1}^{\Nt} \braces{\sum\limits_{r=1}^{\RankC} \lambdaMR \GlLowerBound - \Gammal \sum\limits_{\tilde{r}=1}^{\RankS} \lambdaMRforI  \IlUpperBound} \geq \frac{\commNoisePower \Gammal}{\rhoml}, \forall l.
\end{equation}

So far, we can formulate the transformed problem in the $i$-th iteration for the MA position subproblem as follows
\begin{subequations}
\begin{align}
    \label{pMAInner}
    (\rm P6): \quad & \underset{\qm}{\text{min}} \quad \sum\limits_{n=1}^{\Nt} \left( \partials{f(\qm)}{\xnm}(\xnm - \xnm^{[i-1]}) \right. \notag \\
    & \left. \quad \quad \quad + \partials{f(\qm)}{\ynm}(\ynm - \ynm^{[i-1]}) \right) \\
    \text{s.t.} \quad & \text{(\ref{R1d}), (\ref{R1f}), (\ref{transformedDminConstraint}), (\ref{transformedSINRConstraint}).}\notag
\end{align}
\end{subequations}

It is noticed that the problem (\ref{pMAInner}) is convex, and can be solved efficiently, e.g., using CVX toolboxes~\cite{grant2014cvx}. The optimal solution for (\ref{pMAInner}) is denoted as $\optimalQm$. Then, we can determine the descent direction of $f(\qm)$ as $\optimalQm - \qm^{[i-1]}$ \cite{boyd2004convex}. We then move along $\optimalQm - \qm^{[i-1]}$ with a stepsize $\omega (0 \leq \omega \leq 1)$ to find a minimum $f(\qm)$. Consequently, the result $\qm^{[i]}$ can be expressed as
\begin{equation}
    \label{updatepMA}
    \qm^{[i]} = \qm^{[i-1]} + \optimalOmega (\optimalQm - \qm^{[i-1]}),
\end{equation}
where $\optimalOmega$ is the stepsize that obtains the minimum objective function. 

\subsection{Summary}
By alternately solving (P3) and (P5), we obtain the overall algorithm, as summarized in Algorithm \ref{Alg:MA}. While designing the transmit beamforming and MA positions at the $m$-th slot, the MA position at the $(m-1)$-th slot, i.e., $\bm{q}_{m-1}$, is input to restrain the MAs' moving speed. When $m=1$, $\bm{q}_0$ is set to a UPA with the antenna spacing being half of the wavelength. Then, the MA positions are initialized as $\qm = \bm{q}_{m-1} + \min \bigBraces{\frac{\vmax T}{M}, \Vert \bm{q}_0 - \bm{q}_{m-1} \Vert} \frac{\bm{q}_0 - \bm{q}_{m-1}}{\Vert \bm{q}_0 - \bm{q}_{m-1} \Vert}$. Namely, the initial $\qm$ is as close as possible to $\bm{q}_0$ under the MA moving speed constraint. This expands the search space of the algorithm and avoids getting trapped in local optima. Consequently, subproblems (P3) and (P5) are solved alternately until convergence. After convergence, we use the Gaussian randomization method \cite{luo2010semidefinite} to construct multiple rank-$1$ solutions for $\wcm$ that satisfy all constraints in (P1). The solution that leads to the minimum objective function is finally chosen as the recovered $\wcm$.

Owing to the minimizations of the objective functions in (P3) and (P5), solving the two subproblems iteratively would monotonically decrease the objective function in (P1). Since the total power of the transmit signal is limited, and the movable region of the MAs is bounded, the objective function is lower bounded. Thus, Algorithm \ref{Alg:MA} converges.

As stated in Sec. \ref{subsection:TBF} and Sec. \ref{subsection:pMA}, the subproblems can be solved using the interior point method in the CVX toolboxes. First, the number of the optimized variable for the SDP in (P3) is denoted as $N_\text{var} = 2 \Nt^2 + 9 K$. The number of \emph{linear matrix inequality} (LMI) constraints is $N_{\rm LMI} = K + 2$, contributed by (\ref{LMI1}), (\ref{LMI2}), and (\ref{LMI3}), respectively. Besides, the number of the rows or columns for the matrix of the $j$-th LMI constraint is given by $m_j=7, 1 \leq j \leq K$, and $m_j=\Nt, K+1 \leq j \leq N_{\rm LMI}$. Thus, the complexity of solving the transmit beamforming subproblem is given by $\mathcal{O}(N_\text{var}^2 \sum_{j=1}^{N_{\rm LMI}} m_j^2 + N_\text{var} \sum_{j=1}^{N_{\rm LMI}} m_j^3)$ \cite{9709118}. Subsequently, the worst-case computational complexity for solving (P6) is $\mathcal{O}((\Nt^2/2+4.5\Nt+L)^4\sqrt{2\Nt} \log \frac{1}{\epsilon})$, where $\epsilon$ is the convergence accuracy for the interior point method \cite{luo2010semidefinite}. Thus, the complexity of solving the MA position subproblem is given by $\mathcal{O}(N_\text{I}(\Nt^2/2+4.5\Nt+L)^4\sqrt{2\Nt} \log \frac{1}{\epsilon})$, where $N_\text{I}$ is the iteration number required for the descent direction search method to converge.

\begin{algorithm}[t]
\caption{Proposed algorithm for solving (P1)}
\label{Alg:MA}  
\begin{algorithmic}[1]

\STATE {\textbf{Input: } $\bm{q}_{m-1}$;}
\STATE {\textbf{Initialization: } $\qm$;}
\REPEAT   
   \STATE Given $\qm$, update $\Wcm$ and $\Rsm$ by solving (P3);
    \STATE {$i \leftarrow 0$};
    \STATE {$\qm^{[i]} \leftarrow \qm$};
    \REPEAT
        \STATE $i \leftarrow i + 1$;
        \STATE Given $\Wcm$ and $\Rsm$, obtain $\optimalQm$ via (\ref{pMAInner});
        \STATE Find $\optimalOmega$ that minimizes the average SPEB;
        \STATE Update $\qm^{[i]}$ via (\ref{updatepMA});
    \UNTIL{The decrement of the objective function $f(\qm)$ is below a specified threshold.}
\STATE $\qm \leftarrow \qm^{[i]}$;
\UNTIL{The decrement of the objective function is below a specified threshold.}
\STATE \textbf{Output:} $\Wcm$, $\Rsm$, $\qm$.
\end{algorithmic}
\end{algorithm}  

\section{Numerical Results}
\label{section:Results}

In this section, we conduct comprehensive simulations to validate our joint beamforming and MA position design algorithm. The comparisons with the state-of-the-art method \cite{10478820} are also illustrated. The simulation setup is first introduced, and then the numerical results are presented.
\subsection{Simulation Setup}
In the simulations, we consider an LEO constellation with $K_\text{s}=40$ satellites in each orbital plane. The LEOs' altitude is $H_\text{s}=200 \: {\rm km}$, and the orbital inclination angle $\beta$ is $65^{\rm o}$. The average radius of the earch is $R_\text{e} = 6371 \: {\rm km}$; the constant of gravitation is $G_\text{e}=6.67 \times 10^{-11} \: {\rm m^3/kg/s^2}$; and the mass of the earth is $5.97 \times 10^{24} \: {\rm kg}$. Thus, the considered time interval is $T = T_\text{s} / K_\text{s} \approx 132.59 \: {\rm s}$. This time interval is divided into $M=30$ time slots. The coverage area $\coverageArea$ is defined as a circle centered at $\midBraces{R_\text{e}, 0, 0}^\top$ under GSCS, where the angle between the center of $\coverageArea$ and the area border is $3^{\rm o}$. The longitude and the latitude of the earth is devided into $L_\text{a} \times L_\text{e} = 100 \times 50$ grids, which results in $L=4$ grids within $\coverageArea$. Furthermore, $K=4$ SEs are located at $\midBraces{R_\text{e}, -0.03 \pi, -0.01 \pi}^\top$, $\midBraces{R_\text{e}, -0.01 \pi, -0.03 \pi}^\top$, $\midBraces{R_\text{e}, 0.01 \pi, 0.03 \pi}^\top$, and $\midBraces{R_\text{e}, 0.03 \pi, 0.01 \pi}^\top$, respectively, under GSCS. Note that the trajectory of the LEO satellite, the $L$ grid centers, and the $K$ SEs are all centrosymmetric with respect to $\midBraces{R_\text{e}, 0, 0}^\top$ under GSCS. The carrier frequency is set to $2 \: {\rm GHz}$, corresponding to the wavelength $\lambda=0.15 \: {\rm m}$. The noise powers are $\senNoisePower=\commNoisePower=-110 \: {\rm dBm}, \forall k, l$, at the SEs and imaginary CEs, respectively. The number of the MAs at the LEO satellite and the receiving antennas at the SEs is set to $\Nt=\Nr=16$. The path loss exponent $\gamma=2$. The antenna gains are set as $G_{\rm LEO}=16 \: {\rm dBi}$ and $G_{\rm CE}=3 \: {\rm dBi}$, respectively. $J=1024$ time indexes are adopted when transmitting the signals to guarantee the accuracy of the approximation in (\ref{Rx}). Unless otherwise specified, $\pmax=20 \: {\rm dBW}$, $\Gammal=6 \: {\rm dB}, l=1,\ldots,L$, $\dmin = \lambda/2$, $\vmax=0.075 \: {\rm m/s}$, and $\mathcal{C}=\bigBraces{x,y \mid -1.5 \lambda \leq x \leq 1.5 \lambda, -1.5 \lambda \leq y \leq 1.5 \lambda}$ in the simulations.

\subsection{Numerical Results}

\begin{figure}[!t]
\centering
\includegraphics[width=0.9\linewidth]{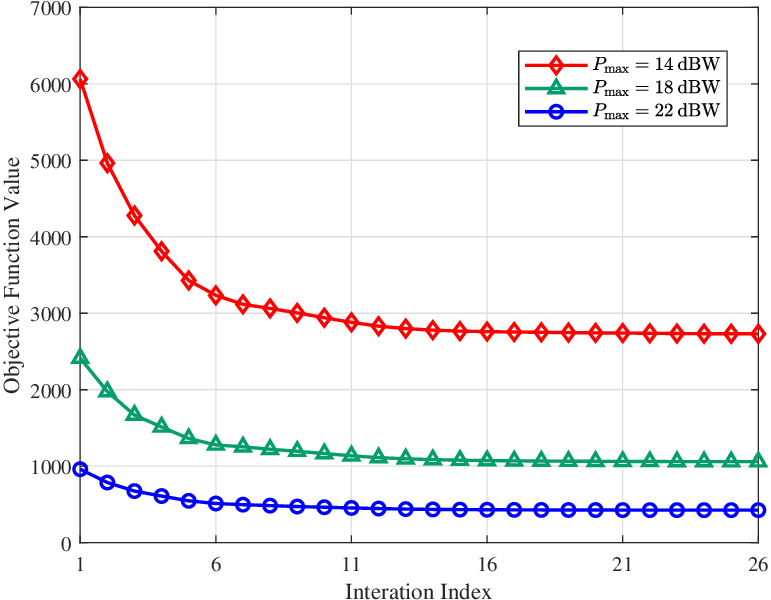}
\caption{The convergence of the proposed Algorithm \ref{Alg:MA} under different settings of $\pmax$.}
\label{convergence}
\end{figure}
We start by assessing the convergence of Algorithm \ref{Alg:MA} in Fig.~\ref{convergence}. Under all settings, the objective function, i.e., the SPEB, monotonically decreases with iterations and typically achieves rapid convergence within $25$ iterations, demonstrating that the algorithm exhibits excellent convergence properties. Moreover, the objective function can converge to lower levels with larger $\pmax$, due to more power and optimization degrees of freedom to achieve better sensing performance while satisfying the communication SINR constraints.

\begin{figure}[t]
\centering
\subfloat[$\Gammal=11 \: {\rm dB}, l=1,\ldots,L$]{\includegraphics[width=0.9\linewidth]{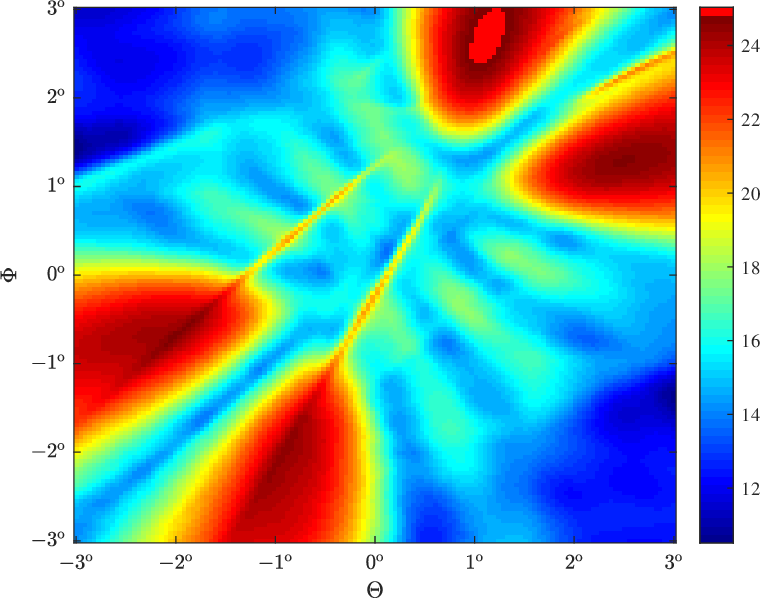}}
\hfill
\subfloat[$\Gammal=20 \: {\rm dB}, l=1,\ldots,L$]{\includegraphics[width=0.9\linewidth]{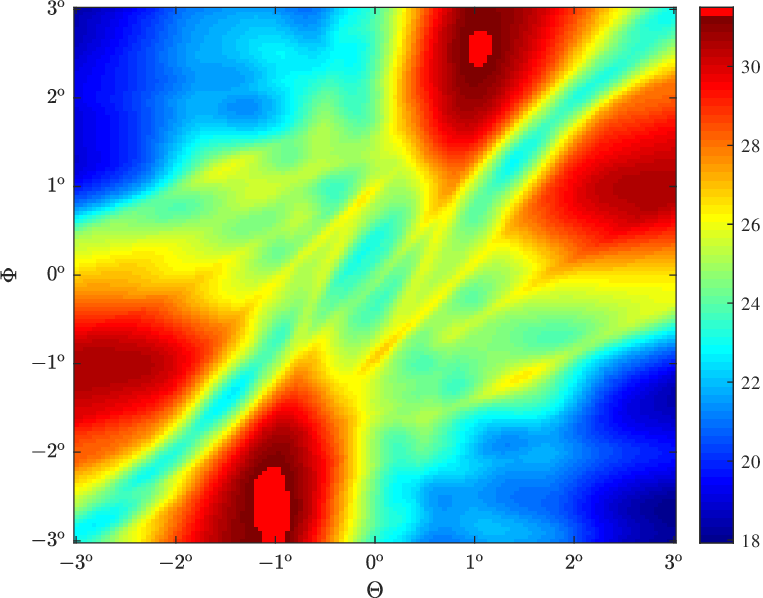}}
\hfil
\caption{The optimized SINR within $\coverageArea$ at different SINR thresholds over $M$ time slots.}
\label{SINR}
\end{figure}

Next, our attention is directed towards evaluating the communication performance. Specifically, we investigate to what extent the method of optimizing the virtual CEs at the $L$ grid centers can enhance the overall communication performance within $\coverageArea$. We first conduct a qualitative analysis. Fig. \ref{SINR} depicts the SINR within $\coverageArea$ at different SINR thresholds, i.e., $\Gammal=11 \: {\rm dB}$ and $\Gammal = 20 \: {\rm dB}$, respectively, where $l=1,\ldots,L$, over $M$ time slots. It reveals that there are $4$ distinct regions where communication performance is markedly prominent. These regions correspond to the vicinity of the $L$ ($L=4$) grid centers selected, which demonstrates the effectiveness of our proposed algorithm. Furthermore, our algorithm not only improves the SINR near the grid centers, but also significantly improves the overall communication performance within $\coverageArea$. It is observed that most areas in $\coverageArea$ meet the required minimum communication SINR.

\begin{figure}[!t]
\centering
\includegraphics[width=0.9\linewidth]{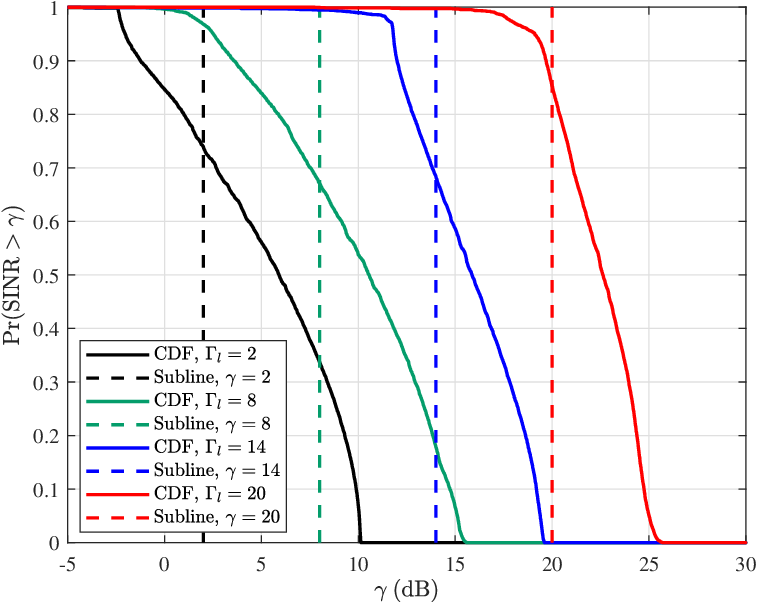}
\caption{The CDF of SINR at different SINR thresholds within $\coverageArea$ over $M$ time slots.}
\label{CDF}
\end{figure}

To further substantiate this finding, we perform a quantitative analysis. Fig. \ref{CDF} depicts the \emph{cumulative distribution functions} (CDFs) of the SINR at different SINR thresholds, where the dashed sublines indicate the thresholds. It is observed that the percentage of satisfaction with the SINR constraint within $\coverageArea$ over $M$ time slots is around $70 \%$ when $\Gammal=2$,  $\Gammal=8$, and $\Gammal=14$. This percentage is achieved even up to $85 \%$ or so when $\Gammal=20$, where $l=1,\ldots,L$. This confirms the conclusions we qualitatively observed from Fig. \ref{SINR}. That is, optimizing the SINR of the $L$ grid centers can significantly enhance the overall communication performance within $\coverageArea$.

\begin{figure}[!t]
\centering
\includegraphics[width=0.9\linewidth]{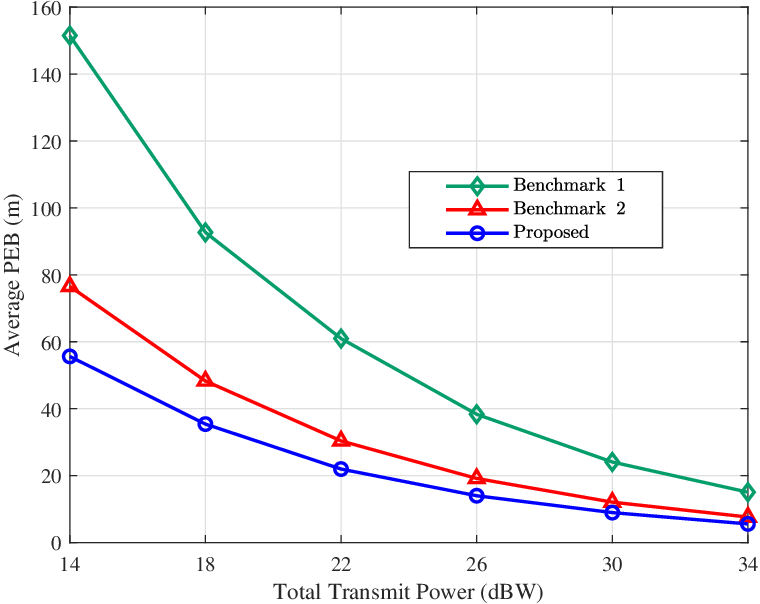}
\caption{Sensing performance comparison of the proposed method with the two Benchmarks at different $\pmax$.}
\label{avgPEBvsPmax}
\end{figure}

Consequently, we compare the sensing performance of our proposed algorithm with \textbf{Benchmark 1: }the random beamforming scheme; and \textbf{Benchmark 2: }the state-of-the-art method proposed in \cite{10478820}, under different total transmit power constraints, i.e., $\pmax$. Benchmark 2 investigates the beamforming design in a massive MIMO-assisted LEO system. An SDR-based optimization algorithm was proposed to minimize the SPEB. Fixed transmit antenna arrays are utilized in both benchmarks, and Benchmark 2 shares the same SINR threshold as the proposed method.

It is observed that the proposed algorithm outperforms Benchmark 1 significantly, even if the SINR constraints are removed due to randomness. Compared to Benchmark 2, our algorithm achieves about $25 \%$ gain in the average PEB, which is achieved by the additional design freedom introduced by MAs. Furthermore, all methods achieve sensing performance gains with the increment of the total transmit power. Under high SNRs, the proposed algorithm can achieve a meter-level localization accuracy.

\begin{figure}[!t]
\centering
\includegraphics[width=0.9\linewidth]{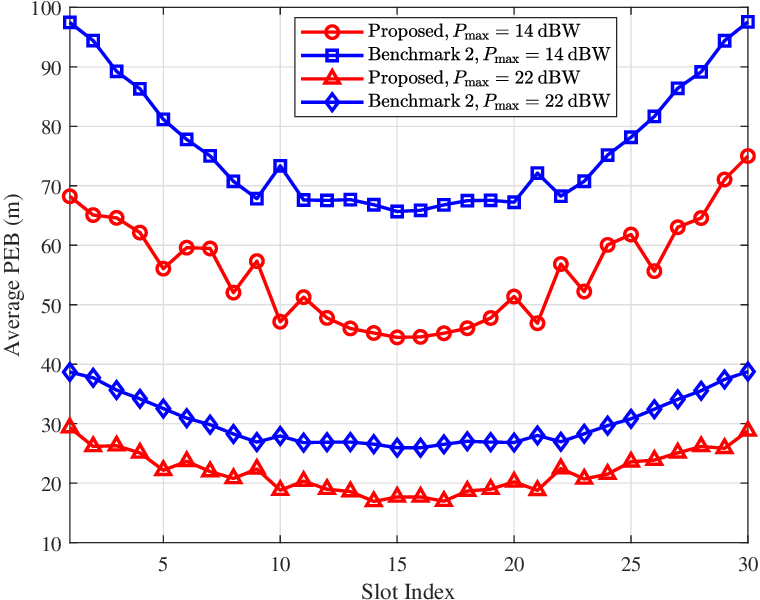}
\caption{Sensing performance comparison of the proposed method with Benchmark 2 under different $\pmax$ during $M$ slots.}
\label{avgPEBvsSlot}
\end{figure}

Further, Fig. \ref{avgPEBvsSlot} compares the sensing performance of the proposed algorithm with Benchmark 2 during all $M$ slots. It is observed that the proposed algorithm effectively enhances sensing performance in each time slot compared to Benchmark 2 in both $\pmax$ settings. Additionally, the sensing performance exhibits symmetry with respect to the intermediate time slot, which stems from the centrosymmetric simulation parameter settings mentioned earlier.

\begin{figure}[!t]
\centering
\includegraphics[width=0.9\linewidth]{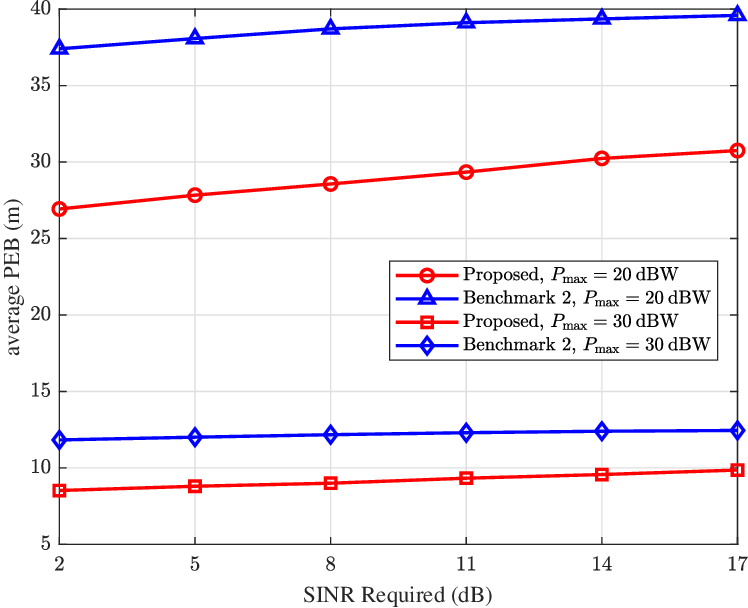}
\caption{The trade-off between communication and sensing performance at different $\pmax$.}
\label{avgPEBvsSINR}
\end{figure}

Finally, Fig. \ref{avgPEBvsSINR} depicts the trade-off between communication and sensing performance at different $\pmax$. It is evident that the proposed algorithm achieves a good trade-off between the two performance metrics since increasing the communication constraints only slightly degrades the sensing performance. Moreover, regardless of the communication constraints and total power budget, our algorithm consistently maintains an improvement of more than $25 \%$ over Benchmark 2, which is consisent with the conclusion drawn in Fig. \ref{avgPEBvsPmax}.

\begin{figure*}[!t]
\centering
\subfloat[$m=1$]{\includegraphics[width=0.25\linewidth]{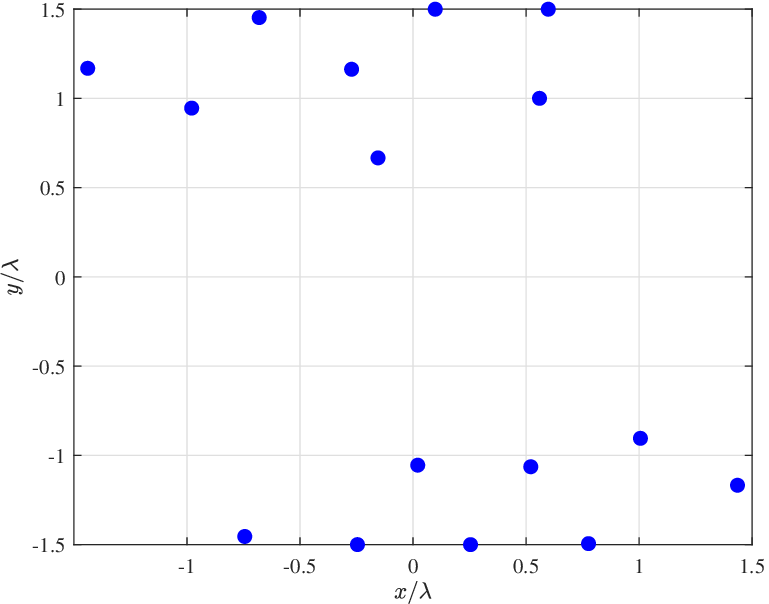}}
\subfloat[$m=5$]{\includegraphics[width=0.25\linewidth]{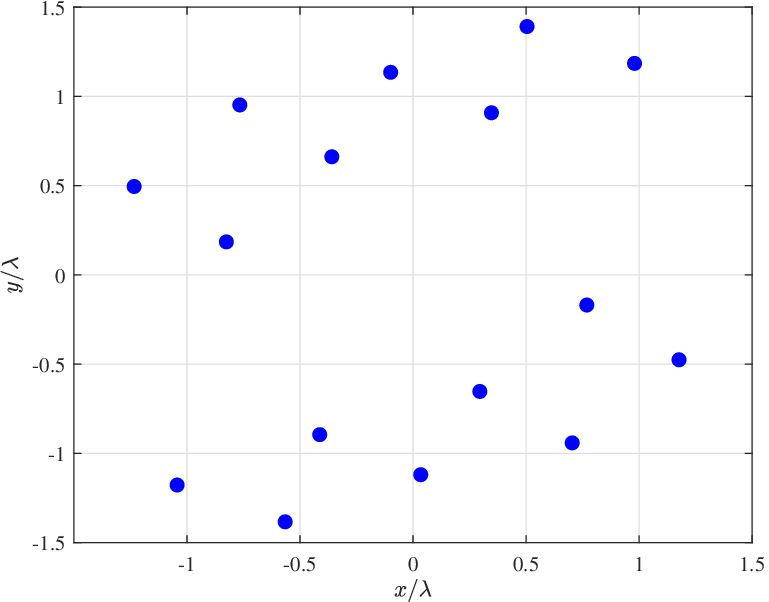}}
\subfloat[$m=9$]{\includegraphics[width=0.25\linewidth]{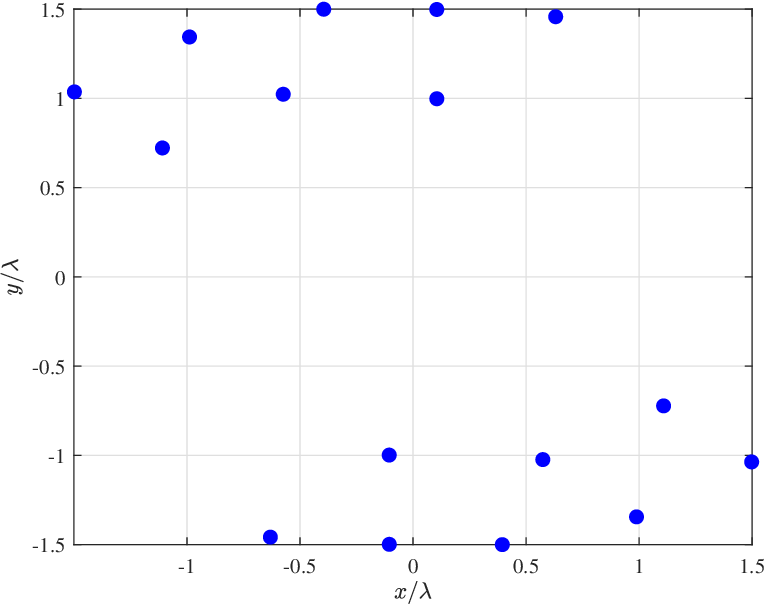}}
\subfloat[$m=13$]{\includegraphics[width=0.25\linewidth]{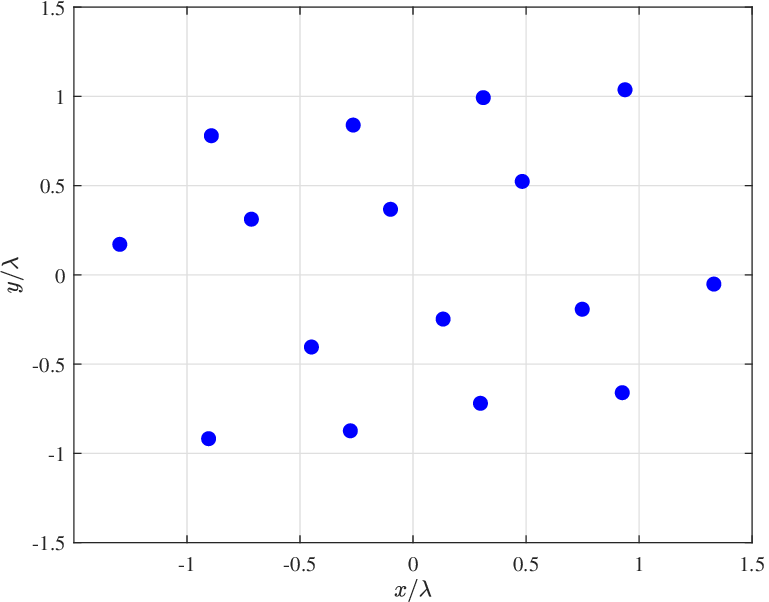}}
\hfil
\subfloat[$m=17$]{\includegraphics[width=0.25\linewidth]{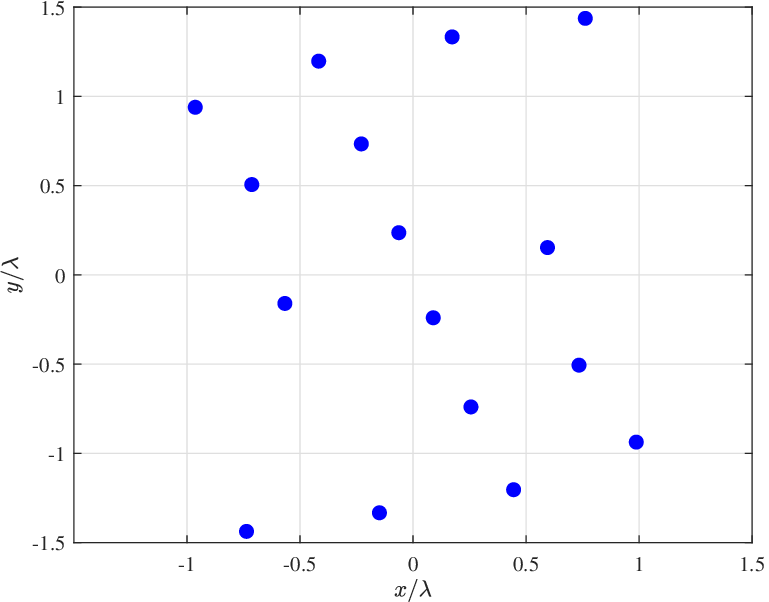}}
\subfloat[$m=21$]{\includegraphics[width=0.25\linewidth]{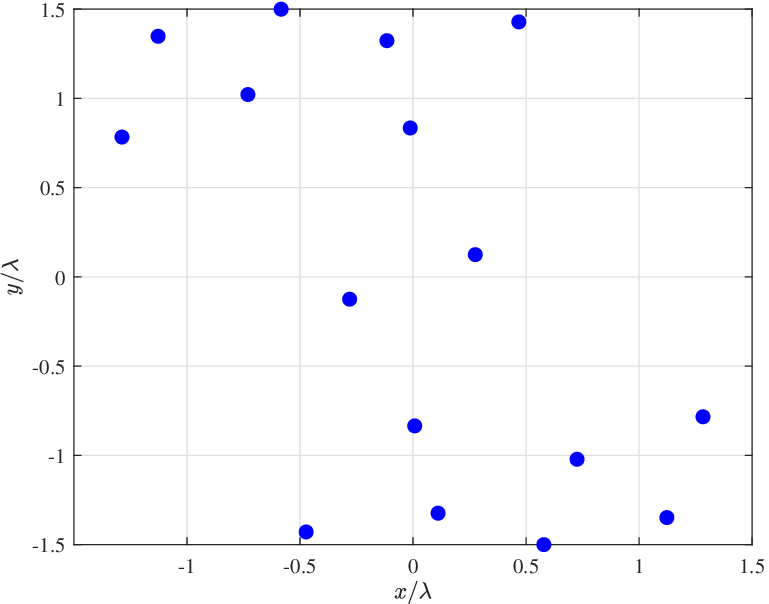}}
\subfloat[$m=25$]{\includegraphics[width=0.25\linewidth]{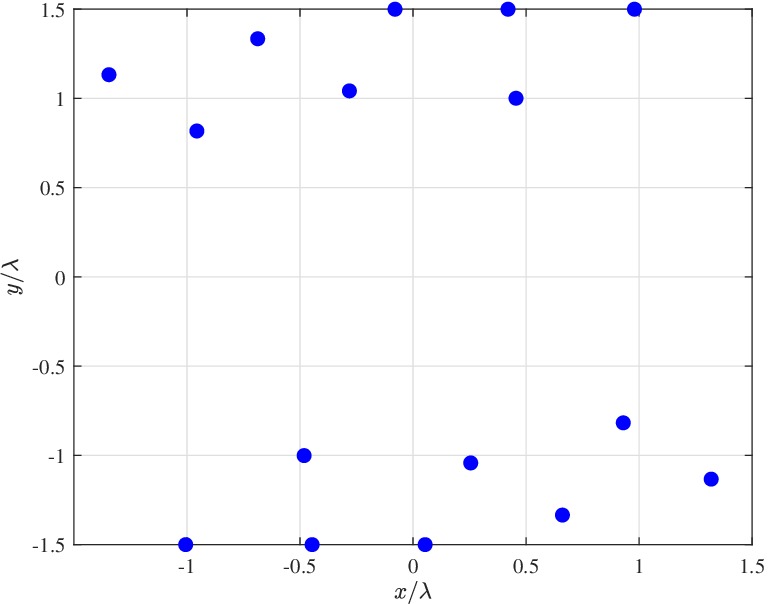}}
\subfloat[$m=29$]{\includegraphics[width=0.25\linewidth]{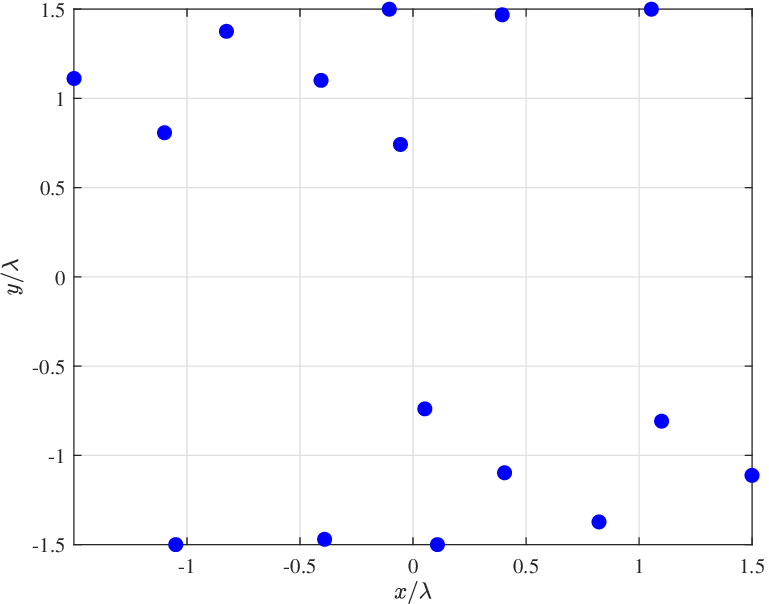}}
\hfil
\caption{The variation of MA positions at different time slots under $\pmax = 20 \: {\rm dBW}$ and $\Gammal=8, l=1,\ldots,L$.}
\label{pMAatDifferentSlots}
\end{figure*}

We also plotted the movement of the MAs when $\pmax=20 \: {\rm dBW}$ and $\Gammal=8, l=1,\ldots,L$, in Fig. \ref{pMAatDifferentSlots}. It is observed that the MA antennas tend to form multiple clusters to assist the LEO ISAC. Within each cluster, the spacing between antennas does not become very large, while the distance between clusters is significantly greater than half of the wavelength. Additionally, the antenna positions also exhibit a symmetrical pattern with respect to the intermediate time slot. This finding cross-validates the conclusion presented in Fig. \ref{avgPEBvsSlot}.

\section{Conclusion}
\label{section:conclusion}
This paper has investigated joint transmit beamforming and MA position design in MA-assisted LEO ISAC systems. An AO-based algorithm based on the SDR has been proposed to minimize the average SPEB under communication SINR, total transmit power, and several MA physical constraints. Numerical simulations have validated the effectiveness of the algorithm. Around $70\%$ tests within the communication coverage area over all time slots meet the SINR constraint. Compared to the benchmarks, our algorithm achieves at least $25\%$ gain in sensing performance. To further improve the performance of the ISAC system, six-dimensional MA can be employed, which additionally adjusts the orientation of the MA array compared to our work, thereby introducing more degrees of freedom for optimization. This will be our future work.

\bibliographystyle{IEEEtran}
{\bibliography{IEEE_REF}}

\end{document}